\documentclass[11pt,a4paper,fleqn]{article}

\usepackage{tabularx, booktabs}
\usepackage{varwidth}
\usepackage{adjustbox}
\newcolumntype{Y}{>{\centering\arraybackslash}X}

\usepackage{datetime}
\usdate

\usepackage[labelfont=bf,textfont=sl,tableposition=top]{caption}

\usepackage{url}
\usepackage{hyperref}
\usepackage{amsmath}

\usepackage{graphicx, color, verbatim, courier}

\usepackage[authoryear,longnamesfirst]{natbib}

\def\tsc#1{\csdef{#1}{\textsc{\lowercase{#1}}\xspace}}
\tsc{WGM}
\tsc{QE}

\usepackage{bbold}

\usepackage{tikz}
\usetikzlibrary{arrows,shapes}
\usetikzlibrary{shapes,decorations,arrows,calc,arrows.meta,fit,positioning}
\tikzset{
-Latex,auto,node distance =1 cm and 1 cm,semithick,
intervention/.style ={rectangle, draw=black},
random/.style ={circle, draw=black},
parameter/.style ={diamond, draw=black},
}

\usepackage{array}
\usepackage{multirow}
\usepackage{authblk}
\usepackage{longtable}
\usepackage{pdflscape}

\usepackage{amsthm, dsfont,bbm}
\usepackage{algorithm, algorithmic}
\usepackage{epstopdf}
\usepackage[autostyle]{csquotes}
\usepackage{color,soul}

\usepackage{marginnote,enumitem,subfigure,rotating}

\newcommand\NECO{\mbox{NECO}}

\newcommand\NECOFest{\widehat{\mbox{NECOF}}}
\newcommand\VaRa{\mbox{VaR}_\alpha}
\newcommand\VaR{\mbox{VaR}}

\definecolor{navy}{rgb}{0,0,0.502}
\definecolor{greenb}{rgb}{0,0.7,0}
\hypersetup{colorlinks,%
citecolor=navy,%
filecolor=green,%
linkcolor=black,%
urlcolor=blue,%
}

\newcommand{\myacknowledgments}{
 \section*{Acknowledgments}
}

\usepackage{setspace}

\begin{document}

\title{Navigating Market Turbulence: Insights from Causal Network Contagion Value at Risk}

\author[1]{Katerina Rigana}
\author[2]{Ernst C. Wit}
\author[3]{Samantha Cook}

\affil[1]{Swiss Finance Institute (SFI), USI Switzerland, Lugano, Switzerland}
\affil[2]{USI Switzerland, Lugano, Switzerland}
\affil[3]{FNA UK, London, UK}

\affil[*]{Corresponding author: katerina.rigana@usi.ch}

\maketitle

\begin{abstract}
Accurately defining, measuring and mitigating risk is a cornerstone of financial risk management, especially in the presence of financial contagion. Traditional correlation-based risk assessment methods often struggle under volatile market conditions, particularly in the face of external shocks, highlighting the need for a more robust and invariant predictive approach.


This paper introduces the Causal Network Contagion Value at Risk (Causal-NECO VaR), a novel methodology that significantly advances causal inference in financial risk analysis. Embracing a causal network framework, this method adeptly captures and analyses volatility and spillover effects, effectively setting it apart from conventional contagion-based VaR models. Causal-NECO VaR's key innovation lies in its ability to derive directional influences among assets from observational data, thereby offering robust risk predictions that remain invariant to market shocks and systemic changes.

A comprehensive simulation study and the application to the Forex market show the robustness of the method. Causal-NECO VaR not only demonstrates predictive accuracy, but also maintains its reliability in unstable financial environments, offering clearer risk assessments even amidst unforeseen market disturbances. This research makes a significant contribution to the field of risk management and financial stability, presenting a causal approach to the computation of VaR. It emphasises the model's superior resilience and invariant predictive power, essential for navigating the complexities of today's ever-evolving financial markets.
\end{abstract}







\section{Introduction}\label{introduction}

Risk is a crucial aspect of financial research, encompassing its definition, measurement, management, and pricing \citep{rubinstein2002retrospective, mcneil2015quantitative}. Whatever its precise definition, there seems to be a consensus that contagion plays a role in the evaluation of financial risks \citep{forbes2001measuring, okimoto2008jfqa,adams2014spillover,caccioli2014stability, glasserman2015likely, dungey2018identifying,loffler2018contagionjfqa, london2019badcontagion, hansen2021financial}.

Causal inference refers to strategies that allow one to draw causal conclusions based on data \citep{pearl2009causality}. This is particularly challenging when experiments are not feasible and the researcher must rely solely on observational data \citep{hill1965environment, warren20142014}. Separating correlation from causality has always been an important issue in any field of empirical research. The increased use of machine learning and AI methods has led not only to a major overhaul of the approach to data handling, but also to a careful reevaluation of causal algorithms for big data \citep{richens2020improving,raita2021big,tchetgen2012causal}. In medicine, this has already led to a renewed evaluation of causal inference methods \citep{gaskell2020introduction, shapiro2021causal,greenland2017and, kroenke2016analysis,williamson2014introduction}. As machine learning and AI methods evolve, more and more research fields are adopting causal inference principles for actionable results, explainability, and safety in applications \citep{amodei2016concrete, pearl2019seven, siebert2023applications, scholkopf2022causality}.
 




The finance industry has been more reluctant to embrace new developments in causal inference, in part as a result of the age-old conundrum of correlation versus causality \citep{embrechts2002correlation, hoover2006causality}. However, there have recently been some examples of using causal inference to improve stress testing \citep{gao2018causal}, empirical research in the accounting field \citep{gow2016causal} and assessing causal factors determining the success of start-ups \citep{garkavenko2022assessing}. 
In finance, the predominant modelling approaches include Granger causality \citep{granger1969investigating,eichler2007granger} and Instrumental Variables \citep{angrist1996identification}. Granger causality, which is part of transfer entropy-based methods \citep{schreiber2000measuring} and applicable to non-linear systems \citep{hlavackova2011equivalence, montalto2014mute}, predicates causality in temporal precedence, assessing if past values of a variable (X) can predict future values of another (Y). In contrast, Instrumental Variables, a facet of quasi-experimental methods \citep{shadish2002experimental}, strive to extract actual causal links from observational data, yet their application hinges on the specific selection of exogenous instruments to address endogeneity. While both models, particularly Granger Causality, hinge on several, often unrealistic assumptions \citep{maziarz2015review, shojaie2022granger}, the usage of Instrumental Variables is seen to be more effective post the establishment of a causal graph through causal inference \citep{sharma2020dowhy}.
Most modelling approaches for the effect of contagion on VaR do not have an explicit causal outset \citep{filho2020role}. Most contagion-based VaR approaches, such as \emph{CoVaR}
\citep{tobias2016covar}, \emph{SDSVaR} \citep{adams2014spillover} and various other alternative attempts to integrate contagion within risk management and the VaR measure
\citep{pesaran2007econometric, metiu2012sovereign, bae2003new}, focus on correlations in systemic risk rather than causality. Furthermore, while insightful, \emph{CoVaR} and \emph{SDSVaR} are complex and often challenging to apply universally. These methods require detailed and extensive datasets and are highly specific, limiting their broad applicability in diverse financial contexts. 

In principle, the use of causal models offers the prospect of robust prediction \citep{peters2016causal,pfister2019invariant, heinze2018invariant}. If cause and effect are correctly identified, then the prediction will offer clear risk guarantees in unexpected circumstances. However, these models are not without limitations. Model uncertainty is an inherent challenge, and under stable financial conditions, simpler predictive methods could outperform causal analysis. However, due to frequent continuous external shocks, financial systems rarely remain in an ergodic equilibrium state, thereby enhancing the benefits of causal models. External influences, such as global events and regulatory changes, underscore the importance of robust and dynamic modelling methods.

In this paper, we apply a causal network approach to risk management. In Section \ref{theory2} we introduce a risk measure that relies on causal network contagion to capture volatility and spillover effects. When assessing the risk position, we consider price changes in related assets that have been identified from the data. Section \ref{sim} analyses the performance of this new approach in a simulation study. Section \ref{forex} applies the method to the Forex market and Section \ref{conclusion2} concludes. 

\section{Value at Risk and Network Contagion}\label{theory2}

\begin{figure}[tb]
	\centering
	\begin{tabular}{cc}
		\includegraphics[width=0.4\textwidth]{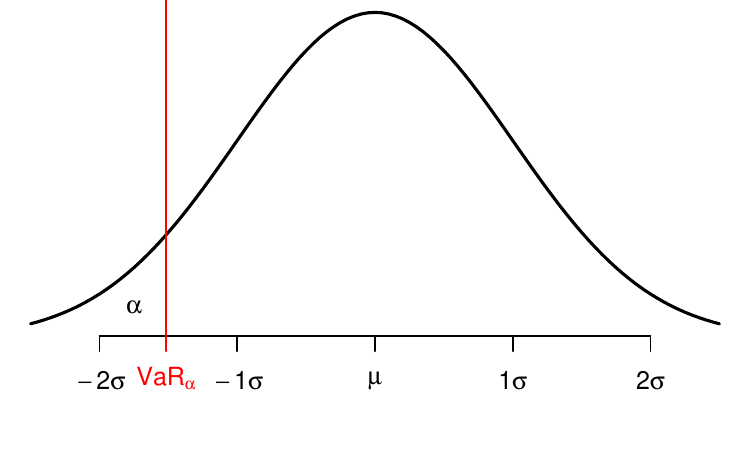}   
		&
		\includegraphics[width=0.4\textwidth]{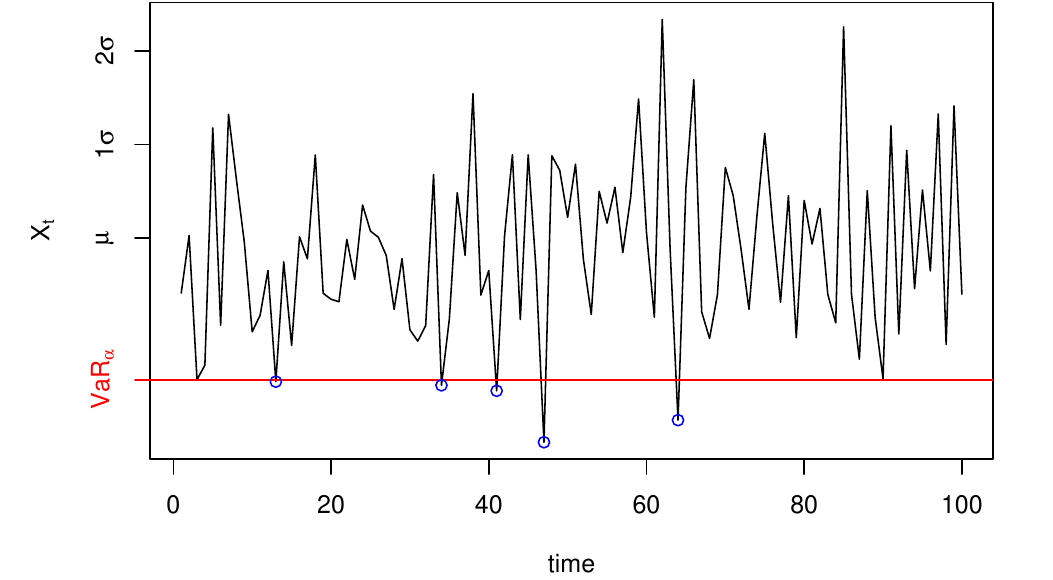}   
		\\
	\end{tabular}
	\caption[Value at Risk]{The Value at Risk, $\VaRa$, is the $\alpha$ quantile of the distribution of a financial instrument. In the example above, the VaR for $\alpha=5\%$ is depicted for a normally distributed instrument both cross-sectionally (left) and longitudinally (right).}\label{fig:VaR-def-normal}
\end{figure}

Value at Risk (VaR) is an established risk measurement used in finance to evaluate and compare the risk of holding a specific financial instrument or portfolio of different instruments. The VaR represents the $\alpha\%$ quantile loss under normal market conditions  
for a specific holding time period \citep{jorion2007value, jorion1996risk2, linsmeier1996risk,dowd2007measuring}. The future value of an asset can be considered a random variable. Identification of the distribution of the value of this instrument $X$ gives access to the value at risk at any level $\alpha$, 
$\VaRa(X) = \inf \lbrace x \in \mathbb{R} : F_X(x) > \alpha \rbrace$, where $F_{X}$ is the cumulative distribution function of X \citep{artzner1999coherent}. This corresponds to the $\alpha$-quantile $q_\alpha(X)$ of the distribution, shown in Figure~\ref{fig:VaR-def-normal}. 
In practical terms, the portfolio value or return will be smaller than the corresponding VaR over the holding period with a probability of at most $\alpha$ per cent. 
%

Usually in finance, the Value at Risk is computed on the basis of log-returns on investment instead of the price for their statistical properties and computational ease. The returns-based VaR is then easily recomposed into the dollar loss VaR using the current price of such a portfolio. 
Value at Risk is a measure developed by investment banks to measure market risk associated with their market positions, a task that is becoming increasingly difficult due to higher volatility periods, increased market interconnectedness, and an increasing use of derivatives. 

Value at Risk (VaR) has evolved from its conceptual beginnings in Edgeworth's 1888 work \citep{edgeworth1888mathematical} to becoming a key financial risk measure. Its development accelerated in the 1980s, culminating in its adoption in the 1990s as the official risk measure for financial institutions \citep{garbade1986assessing, supervision1996amendment}. VaR's ascent was aided by methods such as JP Morgan's \emph{ RiskMetrics}$^{TM}$ \citep{riskmetrics1996riskmetrics}.

There are two types of issues in practice with the VaR. One is the lack of reliable computational methods and the other is a limitation of what it is capable of measuring \citep{artzner1999coherent}. VaR is a measure that is not additive and does not inherently represent the absolute worst-case loss, but is widely used nonetheless. Due to its standardised and easily interpretable nature, it makes an ideal baseline for examining causal network contagion effects in finance. As this paper focusses on VaR, the methodology can be extended to other measures such as expected shortfall (ES), a coherent risk metric that, despite its higher data requirements, can be linked to VaR through \cite{taylor2022forecasting}.

\subsection{Most Common Ways to Estimate Value at Risk}\label{common-VaR}

There are various methods for estimating the VaR. A parametric approach is the variance-covariance method, also called the delta-normal method, popularised by JP Morgan as the RiskMetrics method \citep{riskmetrics1996riskmetrics}. The parametric variance-covariance approach assumes a distribution for the returns on each instrument, whose parameters are then estimated to find the VaR. The most common method is based on the multivariate normal distribution, which leads to an easy and quick estimate of $\VaRa^{\mbox{VC}}(X) =  \mu_x -  z_{\alpha} \cdot \sigma_x$, where $\mu_x$ is the mean and $\sigma_x$ the standard deviation of past realisations of the log-return $X$, and $z_{\alpha}$ is the Z-score of a standard normal distribution at $\alpha\%$. The biggest drawbacks of this method are the, not always realistic, underlying assumptions of normally distributed log-returns, a constant volatility through time, and no correlations among the financial instruments. There is no way to include the effects of contagion from assets not within the considered portfolio in this method, regardless of the distribution assumptions.

To counteract the constant volatility assumption, volatility can also be modelled using ARCH and GARCH models. When dealing with non-linearly priced instruments, such as options, the VaR is usually estimated using a Monte Carlo or stochastic simulation \citep{linsmeier1996risk, linsmeier2000value}. The Monte Carlo VaR is computed as the quantile of the simulated returns $X^{sim}$, simulated from a chosen stochastic model for the behaviour of X -- usually a GARCH process, $\VaRa^{\mbox{GARCH}}(X)=\hat{q}_{\alpha}(X^{sim})$

In non-parametric VaR models, the joint risk factor distribution is constructed using historical data rather than assuming a specific functional form \citep{cabedo2003estimating}. The easiest method would be to resample the past returns within the estimation window and pick the $\alpha$-quantile for the VaR directly. Historical simulation (HS) further develops this idea by bootstrapping past returns, $\VaRa^{\mbox{HS}}=\hat{q}_{\alpha}(X^{\mbox{\scriptsize boot}})$,
where $X^{\mbox{\scriptsize boot}}$ represents the increase in sample through bootstrapping. HS is an easy and fast method. Although it is non-parametric, it still relies on an assumption of stationarity in distribution and specifically volatility, often violated in practice due, for example, to volatility clusterings \citep{gurrola2015filtered}. 

Filtered historical simulation (FHS) combines the best of parametric and non-parametric approaches \citep{barone1998don, barone1999var}. FHS runs an HS method on volatility-rescaled past returns, thus maintaining the non-parametric nature of HS while allowing for varying volatility. Rescaling is done in two steps within a conditional volatility model (e.g., GARCH or AGARCH). Returns are first standardised by the estimated volatility of the day of the return and then rescaled by the forecast volatility for the VaR holding period, $\VaRa^{\mbox{FHS}}(X)=\hat{q}_{\alpha}(X_{re-scaled}^{\mbox{\scriptsize boot}})$. This transformation reflects current market conditions and, therefore, requires shorter estimation windows to simulate extreme events. In contrast to other Monte Carlo based approaches, the correlation matrix does not have to be estimated, as all rescaled returns at a specific event time are sampled together. Extreme observations or correlations that vary over time may not be adequately considered by the FHS VaR \citep{pritsker2006hidden}. Extensions of the FHS have been proposed by \cite{hull1998incorporating} as the Volatility-weighted HS 
and by \cite{mcneil2000estimation}, who combine FHS with extreme value theory. For comparisons, see \cite{dowd2007measuring}, \cite{martins2018nonparametric}, and \cite{pritsker2006hidden}.

\subsection{Definition of Causal NECO Value at Risk}
\label{sec:necovar}

In this section, we propose a novel VaR procedure based on causal network contagion (causal NECO). 
What separates this approach from standard dependency analysis like \cite{jondeau2006copula}, is the ability to identify the direction of the contagion: which assets export contagion risk and which others import it? 
The method does not assume any a priori causal model and considers only what can be gathered through observed data --- through the external manifestation through the path of returns for each asset and their co-dependences. Our goal is to demonstrate how causal contagion from assets outside the portfolio of interest can be identified from observational data. 

\paragraph{Gaussian Causal NECO VaR.}
In this paragraph, we first derive the form for the value at risk in the event that the causal contagion system can be described by Gaussian noise. We assume that the log-returns for each instrument i at time t ($X_{i,t}$) are described by a structural equation model (SEM), consisting of an autoregressive part and a contagion part, 

\begin{equation} \label{eqn:commented}
X_{i,t} \leftarrow 
 \alpha_{0,i}+ \sum_{\ell=1}^{L} \alpha^{\ell}_{i}X_{i,t-l}
+ \sum_{j\in pa(i)}\beta_{ji}X_{j,t}+\varepsilon_{i,t} \quad 
\quad \varepsilon_{i,t}\sim N(0,\sigma_i^2) 
\end{equation}
\normalsize
where $L$ is the number of lags considered for the autoregressive part, $\alpha_0$ is the intercept, $\alpha^{\ell}_{i}$ are the autoregressive coefficients of lag $\ell$ for asset $i$, $pa(i)$ are the causal parents of instrument $i$,  $\beta_{ji}$ are the causal effects of instrument $j$ on instrument $i$, and $\varepsilon_{i,t}$ are the error terms, assumed to be independent and normally distributed with variance $\sigma^2$. Here, contagion is defined as the instantaneous causal structure of the financial system according to \cite{rigana2021causal}, which also provides more information on the choice of assets. The SEM can be written more concisely in matrix form for all financial instruments in the causal contagion network:
\begin{equation}\label{eqn:matrixversion}
X_{t}=(\mathbb{1}-B)^{-1}\left(\alpha_0 +A \cdot X_{t-1:t-\ell}+ \varepsilon_{t}\right)
\end{equation}
where the $\varepsilon_t\sim N(0,\Sigma)$ is the vector of noise components, $A$ is the matrix of the autoregressive coefficients and $B$ the matrix of contagion effects,
\begin{equation}
A = 
\begin{bmatrix}
\alpha_{11} &\ldots & \alpha_{1\ell}\\ 
\alpha_{21} &\ldots & \alpha_{2\ell}\\ 
&\cdots& \\ 
\alpha_{p1} &\ldots & \alpha_{p\ell}
\end{bmatrix}, \quad
B = 
\begin{bmatrix}
0 & \beta_{12} & \cdots & \beta_{1p} \\
\beta_{21} & 0 & \cdots & \beta_{2p} \\
\vdots & \vdots & \ddots & \vdots \\
\beta_{p1} & \beta_{p2} & \cdots & 0
\end{bmatrix}, \quad
\Sigma = 
\begin{bmatrix}
\sigma_1^2 & 0 & \ldots & 0 \\
0 & \sigma_2^2 & \ldots &0 \\
0 & 0 & \ddots &0 \\
0 & \ldots & & \sigma_p^2
\end{bmatrix} 
\end{equation}\label{eqn:details}
Though resembling a traditional vector autoregressive model, this SEM model is strictly causal, where $\beta_{i,j}$ shows the direct causal effect of $i$ on $j$. The identifiability of this causal model is explained in Section \ref{estimation}.

Given Equation \ref{eqn:matrixversion}, the distribution of our $X_t$, conditional on the past log-returns, is given as:
\begin{equation} \label{eqn:distribution}
X_{t}|X_{t-1:t-\ell} \sim N\left((\mathbbm{1}-B)^{-1}\left(\alpha_0+A \cdot X_{t-1:t-\ell}\right), 
(\mathbbm{1}-B)^{-1} \Sigma(\mathbbm{1}-B)^{- \top}\right) 
\end{equation}
The contagion matrix $B$ has a direct impact on both mean and volatility. This is not just a theoretical point; it shows that contagion really does have a practical impact on these critical financial metrics.

Using the distribution in Equation \ref{eqn:distribution}, we can incorporate causal network contagion (NECO) into the VaR definition. The Gaussian Causal VaR for each instrument $i$ corresponding to a risk level of $\alpha$
\begin{equation} \label{eqn:neco-var}
\VaRa^{\mbox{NECO}}(X_{(t,i)})=\left[\begin{array}{c}(\mathbbm{1}-B)^{-1}\left(\alpha_0+A X_{t-1:t-\ell}\right)\end{array}\right]_i - z_{\alpha} \sqrt{\left[(\mathbbm{1}-B)^{-1} \Sigma(\mathbbm{1}-B)^{- \top}\right]_{i i}}
\end{equation}
\paragraph{General Causal NECO VaR.}
The definition of the Causal VaR above is based on the assumption of a multivariate normal distribution for the instruments considered for the causal networks. This assumption represents log-returns relatively well, but it is still often violated in reality where financial instruments are concerned. 
In this paragraph, we will relax the normality assumption by using the Gaussian copula transformation to capture fat tails and other non-Gaussian behaviour typical of the returns on financial instruments \citep{dobra2009copula,abegaz2015copula}. Copulas provide a flexible tool for understanding the dependence between random variables, particularly for non-Gaussian multivariate data \citep{mohammadi2017bayesian}.

Given that logarithmic returns of the instrument $i$ follow a marginal distribution $F_i$, we assume that there is a latent temporal process $Z$, which can be described by the mean causal SEM in Equation \ref{eqn:matrixversion}. This SEM describes the idealised Gaussian version of the financial process, centred around zero with variance one. The connection between the idealised process $Z$ and the observable $X$ is given by the copula transformation, 
\begin{equation}
	X_{t,i}=F_{i}^{-1}\left(\Phi\left(Z_{t,i} \right)\right),
\end{equation}
where $\Phi$ is the CDF of a standard normal distribution. This gives us a direct way to define a general causal VaR,
\small 
\begin{equation} \label{eqn:neco-var-general}
	\VaRa^{\mbox{NECO}}(X_{(t,i)})=F_i^{-1}\left( \Phi \left(\left[(\mathbbm{1}-B)^{-1}A Z_{t-1:t-\ell}\right]_i - z_{\alpha} \sqrt{\left[(\mathbbm{1}-B)^{-1} (\mathbbm{1}-B)^{- \top}\right]_{i i}}\right)\right),
\end{equation}
\normalsize
where $Z_{t,i} = \Phi^{-1}(F_i(X_{t,i}))$. 
The copula transformation allows us to perform our analysis in the latent space without the need to assume lognormal distribution for the returns. The other advantage of using a Gaussian copula is that, unlike most other copulas, it can handle high dimensionality, i.e., many interrelated financial instruments. Gaussian copulas gained a negative reputation due to their indiscriminate use during the 2008 financial crisis, particularly in modelling complex financial instruments whose risks were poorly understood. However, it is essential to distinguish between the misuse of these tools and their inherent capabilities. In our context, we do not employ the Gaussian copula to model risk, we use them to enhance the process of estimating causal networks as shown in the following section.

\subsection{Estimation of Causal $\NECO$ Value at Risk} \label{estimation}

Given a multivariate time-series, $D_X=\left\{ x_{t,i}  \right\}_{it}$, of log-returns of $p$ financial instruments across $N$ periods, the estimation of the Causal $\NECO$ VaR is done in four steps. First, we estimate the underlying marginal distributions $F_i$ for the instruments. Second, we estimate the causal structure on the transformed scale to see what financial instruments impact the risk of the instrument of interest. This pivotal step is instrumental for transitioning from mere correlation analysis to a causal model. Thirdly, given the causal structure, we then estimate the contagion coefficients $A$ and $B$ from Equation \ref{eqn:details}. Finally, the estimated marginal distributions $\widehat{F}_i$ and the coefficients $\widehat{A}$ and $\widehat{B}$ are used to calculate the VaR as in Equation \ref{eqn:neco-var-general}.

\paragraph{Step 1: Estimating the marginal distribution of financial instruments} For each instrument $i$, we estimate its marginal distribution $F_i$ non-parametrically as the adjusted empirical distribution function,
\[ \widehat{F}_i(x) = \frac{0.5+ \sum_{i=1}^N 1_{\left\{ x_i\leq x\right\}}}{N+1}. \] 
We then use this empirical distribution, to define a transformed dataset, $D_Z$, of normally distributed variables,
$D_Z = \left\{z_t\in \mathbbm{R}^p~|~z_{t,i} = \Phi^{-1}(\widehat{F}_i(x_{t,i})), t=1,\ldots,N \right\}.$
We use the adjusted empirical distributions, in order to avoid degenerate values for $z$. 

\paragraph{Step 2: Discovery of Causal Structure} For the discovery of the causal structure, we follow \cite{rigana2021causal}. We estimate the causal structure based on transformed data $D_Z$ in the form of a causal network. This method is based on the PC-stable algorithm of \cite{colombo2014order}. The PC-stable algorithm is a more robust version of the original PC algorithm \citep{spirtes1991algorithm}. We implement the PC-stable algorithm with the R package {\tt pcalg} \citep{kalisch2012causal, pcalg2hauser}.  

The fundamental insight of the PC algorithm is that causal connections in the structural equation model can be estimated from purely observational data. The algorithm is based on two main insights. \emph{Conditional independence}: if two variables are independent or conditionally independent given a third variable, then they are not directly causally connected. \emph{Collider}: if two variables are (conditionally) independent, but become dependent when conditioning on a third variable, then the original two variables are the causal parents of the third. Using these insights, we can reconstruct the complex causal dynamics within a system in terms of a causal network and identify direct and indirect relationships.


%
%
%
%
%
%
%
%
%
%

\paragraph{Step 3: Estimation of Causal Effects}
Once the causal structure of the SEM in Equation \ref{eqn:commented} is established, the coefficients of matrices A and B can be estimated using standard least squares. If there are no links present in the causal network between the variable $i$ and $j$, then the corresponding coefficients $\beta_{ij}$ and $\beta_{ji}$ are set to zero. If the algorithm from the previous step is unable to deliver a fully directed network --- where for some couples of instruments both directions of the contagion are just as likely given the structure of the causal network --- we obtain a multiset of possible coefficients and combine these into a range estimator \citep{maathuis2009estimating}. Given that the financial system is very big and interconnected, the chances of finding a fully directed network are very high.

\paragraph{Step 4: Causal NECO Value at Risk}With the estimated $\widehat{F}_i$, $\widehat{A}$ and $\widehat{B}$ we can then compute the VaR using Equation \ref{eqn:neco-var-general}. We assume that any interim payments on the considered assets are either zero or reinvested continuously in the asset itself, as is done in mutual funds. Furthermore, we take the time period for the VaR evaluation to be equal to the frequency of the recorded log-returns --- so if we consider daily log-returns, we will compute the 1-day $\VaRa$.

\section{Performance of Causal NECO Value at Risk}\label{sim}

In this section, we conduct a detailed analysis of the Causal NECO Value at Risk (VaR) in its ability to accurately predict risk levels in a variety of simulation studies. The efficacy of Causal NECO VaR must be juxtaposed with four other established methods, as delineated in Section \ref{common-VaR}. We introduce the backtest techniques used in Section \ref{compare-def}. We pick backtesting measures that are model-free and best suited for comparative analysis.


\subsection{Backtesting Measures} \label{compare-def}

Using simulations or historical data, backtesting can be used to assess how well VaR performed. \cite{campbell2005review} provides an overview of the most common backtesting methods. Part of the backtesting will be performed using the R package {\tt MSGARCH} \citep{ardia2016generalized}. The following is a description of the tests we will consider. 

\paragraph{Average Exceedance Rate.} The computationally easiest and most immediate test for VaR performance is looking at the number of times that the actual $X_t$ log-returns fell below the VaR, called violations or exceedances. Given the $\VaR$ definition, we expect $X_t$ to be smaller than $\VaRa$ approximately $\alpha \%$ times. The average exceedance rate $\hat{\alpha}$ should therefore be close to the target level $\alpha$ of $\VaRa$.

\paragraph{Actual over Expected Ratio.} Actual over Expected Ratio (AE) measures whether the VaR computation method tends to have more or fewer violations than expected given the target $\alpha$. A good method has the number of exceedances $n_1$ being $\alpha  N$. Defining $AE=\frac{n_{1}}{\alpha  N}$, $AE>1$ means that the method is not restrictive enough and underestimates the risk of the underlying investment. An $AE<1$ shows the opposite, a method that is overly conservative and overestimates the risk. Both directions of this error can lead to costly mistakes.

\paragraph{LR Test of Unconditional Coverage.} The Likelihood Ratio test (LR) assesses the coverage rate for the target rate $\alpha$. This unconditional coverage test (UC), as proposed by \cite{kupiec1995techniques}, evaluates whether the observed proportion of violations deviates significantly from the expected level. The LR test statistic, defined as $LR_{UC} =-2 ln \left[\frac{(1-\alpha)^{n_0} \cdot \alpha^{n_1}}{(1-\hat{\alpha})^{n_0} \cdot \hat{\alpha}^{n_1} } \right]$, follows an asymptotic $\chi^2$ distribution under the null hypothesis $H_0: E \left[ \hat\alpha  \right]  = \alpha$.

\paragraph{LR Test of Conditional Coverage.} The Likelihood Ratio test (LR) test for the Conditional Coverage (CC) from \cite{christoffersen1998evaluating} expands the UC test for the detection of violations clusterings in time. No violation occurrence should be informative about the performance of the next-step VaR. CC checks that exceedance realisations $\{ \mathbbm{1}_{\{ x_{1,i} \leq \mbox{\scriptsize VaR}_\alpha(1,i)\} } , \cdots  , \mathbbm{1}_{\{ x_{T,i} \leq \mbox{\scriptsize VaR}_\alpha(T,i)\} } \}$, often called hit series, are distributed independently and identically. It uses a likelihood ratio test with 2 degrees of freedom. 

\paragraph{Dynamic Quantile Test.} The Dynamic Quantile (DQ) test described in \cite{engle2004caviar} and \cite{dumitrescu2012backtesting} tests whether the violations, i.e. exceedance realisations, are not only uncorrelated among themselves, but also with other lagged variables.
The $LR_{DQ}$ follows asymptotically a $\chi^2_p$ distribution with $p$ degrees of freedom.
\paragraph{Absolute Deviation.} The mean and maximum absolute deviation (AD) show the actual loss that would occur if an investor or bank had relied on the VaR prediction. As pointed out by \cite{mcaleer2008forecasting}, the AD measure is of great importance as large violations can lead to bank failures when the capital requirements implied by the VaR threshold forecasts are not sufficient to protect against losses that are actually realised.
\paragraph{Average Quantile Loss.} 
The Average Quantile Loss \citep{gonzalez2004forecasting} is a weighted loss measure, defined as 
$QL_{i}(\alpha) = \frac{\sum_{t=1}^N \left(\alpha-\mathbb{1}_{\{ x_{t,i} \leq \mbox{\scriptsize VaR}_\alpha(t,i)\} }\right)\left(x_{i,t}-VaR_{\alpha}(t,i)\right)}{N}.$
If for two methods we have $ QL^{1} <  QL^{2}$,  then method 1 is preferable over method 2. 


\subsection{Comparison of NECO VaR with Other Methods}
\label{sec:compmethods}
\begin{figure}
	\begin{tabular}[p]{cc}
		\begin{minipage}[p]{0.6\textwidth}
\begin{tikzpicture}	
	\node[random] (x5) at (0,0) {$X_{5}$};
	\node[random] (x3) at (2.5,1) {$X_{3}$};
	\node[random] (x4) at (5,0) {$X_{4}$};
	\node[random] (x2) at (2.5,3) {$X_{2}$};
	\node[random] (x1) at (7.5,0){$X_{1}$};
	
	\path[draw,thick,->]
	(x1) edge (x4)
	(x2) edge(x3)
	(x2) edge(x4)
	(x2) edge (x5)
	(x3) edge (x4)
	(x3) edge (x5)
	(x4) edge (x5);
\end{tikzpicture}
		\end{minipage}
&
\begin{minipage}[p]{0.3\textwidth}
\begin{tabular}{rr}
	\hline
	\multicolumn{2}{c}{Simulated Network}\\ 
	\hline
	nr. links & 7 \\ 
	density & 0.7 \\ 
	market NECOF & $47\%$ \\ 
	\hline
\end{tabular}
\end{minipage}
\end{tabular}
\caption{Simulated financial network with 5 instruments related to section~\ref{sec:compmethods}}
\label{fig:causal-graph1}
\end{figure}
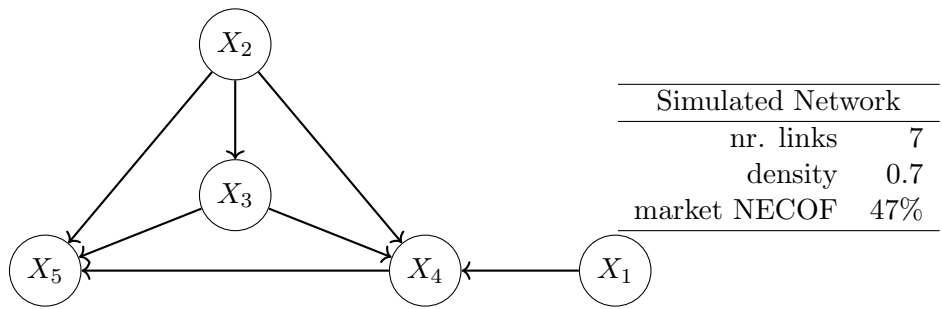

We start by comparing the performance of our causal VaR method with more traditional methods. We simulate data for the contagion network of 5 financial instruments, as shown in Figure \ref{fig:causal-graph1}. We choose a relatively difficult market situation with a relatively high density and market contagion to simulate a market in crisis. The contagion coefficients were randomly chosen to reach a market contagion of 47\%. Using the Network Contagion Factor (NECOF) measure, we can target a particular market contagion. NECOF expresses the impact of contagion in \%, where 0\% means that no contagion effects are detected and 100\% means that all changes can be explained by contagion alone \citep{rigana2021causal}. We examine the ability of various methods to deal with non-normality using exponentially distributed returns with an added shock every 100 days. Out-of-sample performance is tested for 100 days after training on 250 time points. The process is repeated twenty times.

\begin{figure}[tb]
	\begin{tabular}{cc}
\includegraphics[width=0.45\textwidth]{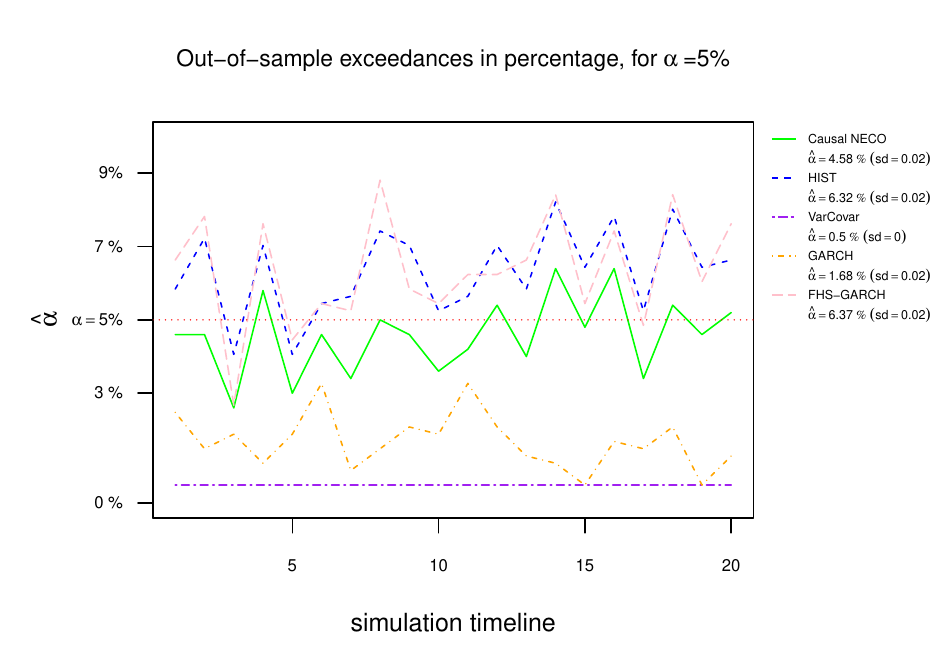}
&
\includegraphics[width=0.5\textwidth]{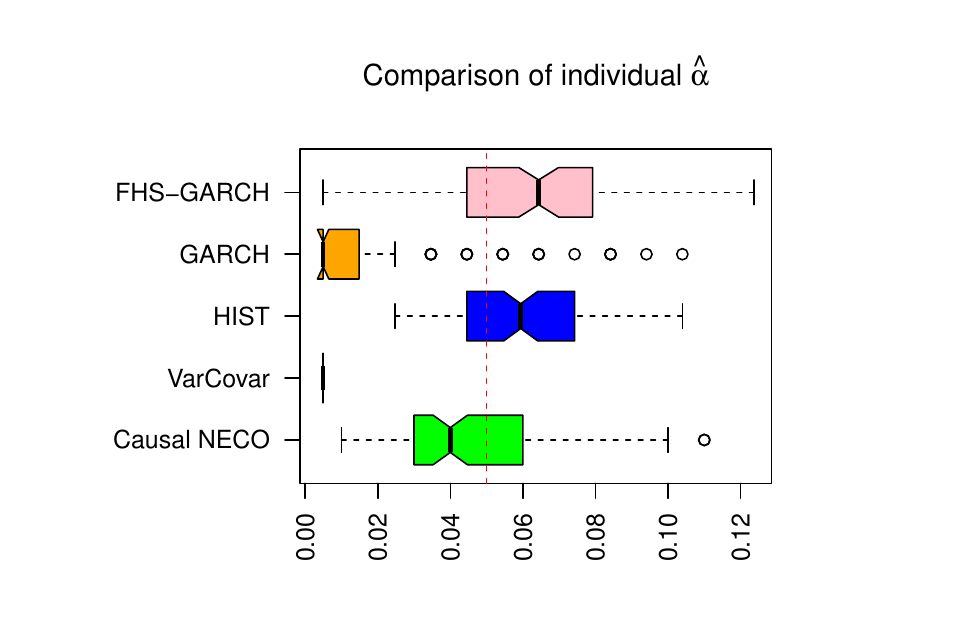}
\\
(a) & (b)
	\end{tabular}
\caption{Fraction of VaR exceedances within the out-of-sample window of 100 days for each Value-at-Risk model for all of the considered financial instruments. Figure (a) illustrates how closely it follows the target of $5\%$ in different simulations, and (b) shows the overall precision with a boxplot.}
\label{fig:sim1_out}
\end{figure}

Figure \ref{fig:sim1_out} shows both the temporal performance and the overall performance of the various VaR methods targeting $\alpha=0.05$. It shows that the causal NECO VaR outperforms the other methods, both in its ability to deal with the non-normality and the external shocks to the system. Table \ref{tab:backtestsim} shows that the causal NECO method has good coverage for the three tests, $LR_{UC}$, $LR_{CC}$ and $LR_{DQ}$, achieving high acceptance rates of 94\%, 96\% and 88\%, respectively.
From Figure \ref{fig:sim1_out} it is immediately clear that the two methods that struggle the most are VarCovar and GARCH; these two methods rely heavily on the assumption of normality. 
FHS-GARCH significantly improves the GARCH method, but since it is based on merging HIST and GARCH, it is influenced by shocks as much as HIST in underestimating the risk.

\begin{table}[tb]
\centering
\begin{tabular}{rrrrrr}
  \hline
 & Causal-NECO & VarCovar & HIST & GARCH & FHS \\ 
  \hline
   mean$(\hat{\alpha})$  & {\bf 0.0458} & 0.0050 & 0.0632 & 0.0168 & 0.0637 \\ 
   st.dev$(\hat{\alpha})$ & 0.0203 & {\bf 0} & 0.0196 & 0.0236 & 0.0257 \\ 
  LRuc.accept &\textbf{ 0.9400} & 0 & \textbf{0.9400} & 0.2100 & 0.8600 \\ 
  LRcc.accept & \textbf{0.9600 }& 0 & 0.9300 & 0.2900 & 0.8200 \\ 
  DQ.accept &\textbf{ 0.8800 }& 1.0000 & 0.7400 & \textbf{0.8800} & 0.6200 \\ 
   AE.mean & \textbf{0.9160} & 0.0000 & 1.1760 & 0.2400 & 1.1860 \\ 
  AE.sd & 0.4052 & \textbf{0.0000} & 0.3962 & 0.4774 & 0.5182 \\ 
  AD.mean &\textbf{ 0.0303} & NA & 0.0375 & 0.1249 & 0.0634 \\ 
  AD.max & \textbf{0.1200} &  -Inf & 0.1300 & 1.9500 & 2.2600 \\ 
  CompareQL & 1.0000 & 2.5811 & \textbf{0.9997} & 1.9885 & 1.0329 \\ 
   \hline
\end{tabular}
\caption{Simulation Backtesting Results, for $\alpha=5\%$. The best results for each measure are indicated in bold. The results reflect all 20 rounds of 100 out-of-sample predicted VaR.}
\label{tab:backtestsim}
\end{table}


%
%
%

\begin{figure}[]
    \begin{tabular}{cc}
\includegraphics[width=0.45\textwidth]{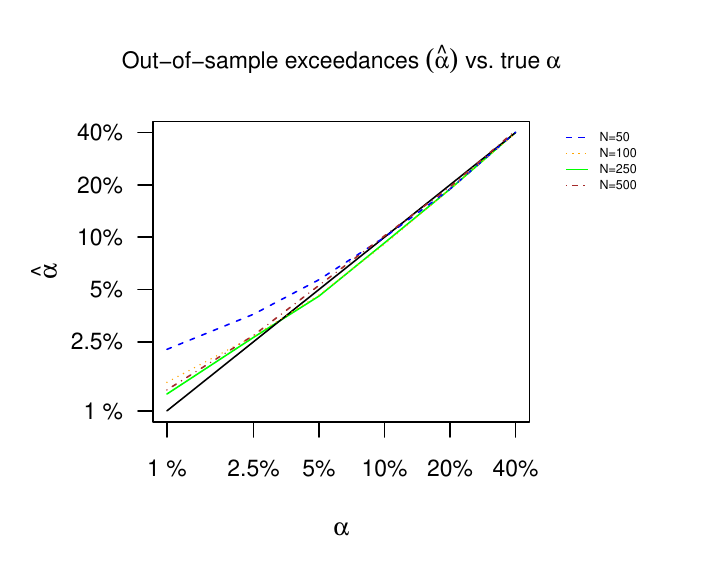}   
&
\includegraphics[width=0.45\textwidth]{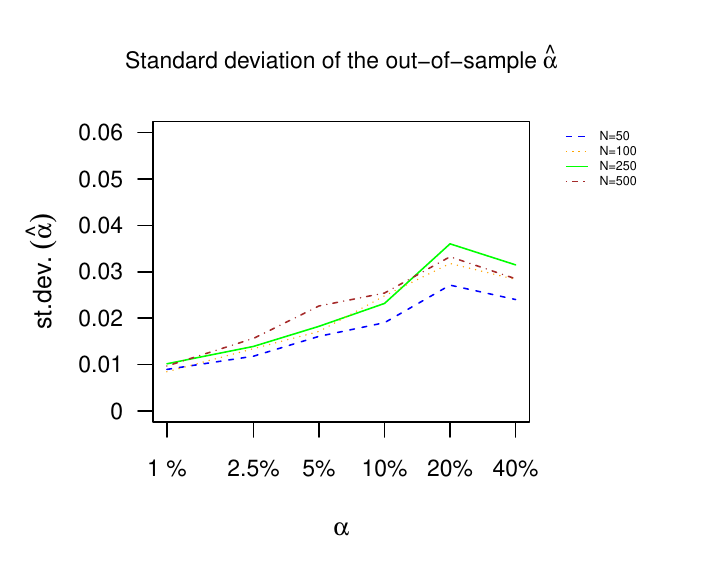}   
\\
(a) & (b)\\
\includegraphics[width=0.45\textwidth]{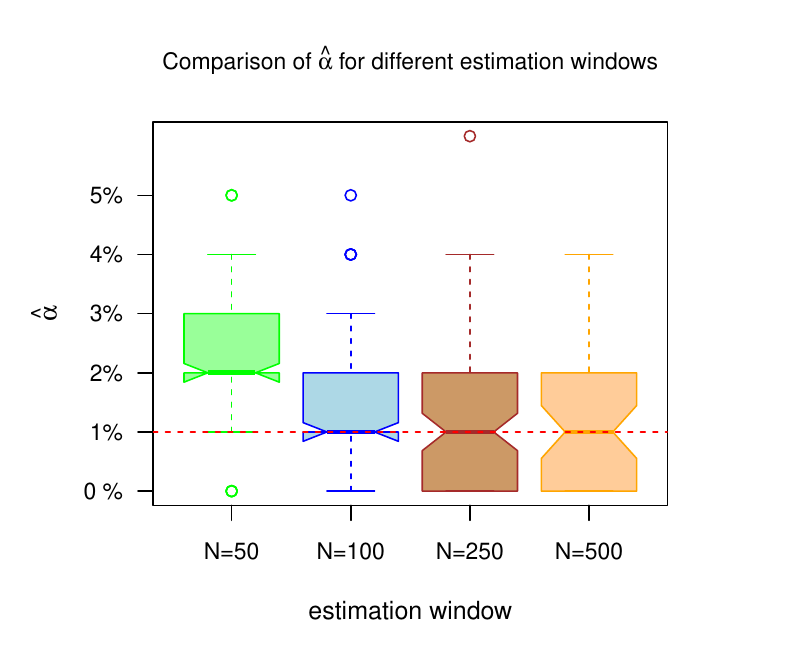} 
&
\includegraphics[width=0.45\textwidth]{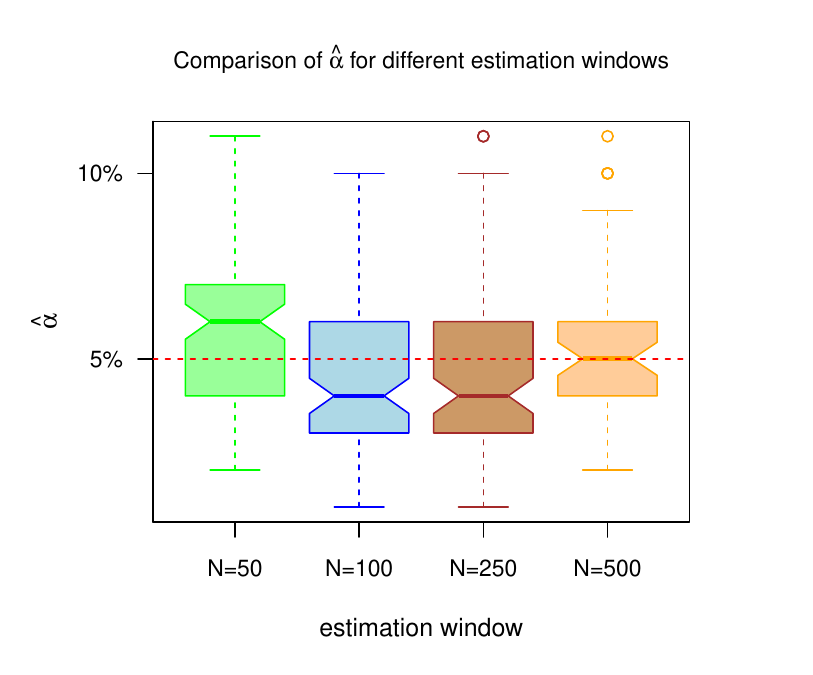} \\
(c) & (d)
\end{tabular}
\caption{Performance of the Causal NECO VaR for different estimation windows. (a) Comparison on the logarithmic scale between different target $\alpha$ levels and different lengths of the estimation window in terms of the considered number of observations (N). The black line in the centre is the target to be achieved. (b) Standard deviation of the individual $\hat{\alpha}$. (c) Comparison at target $\alpha=1\%$ for different N. (d) Comparison at target $\alpha=5\%$ for different N.}
\label{fig:sim-window}
\end{figure}

\subsection{Effect of Training Window}

The performance of any VaR method depends on the accuracy of the estimated model. In this simulation study, we vary the training window from $N=50$ to $N=500$ observations to fit the causal model. Once the causal structure is estimated, we apply the causal NECO VaR to an out-of-sample time series of 100 time points. Figure~\ref{fig:sim-window} shows the results. Although the standard deviation of the achieved VaR level does not depend much on the size of the training window $N$ as seen in
Figure 4 (b), the VaR tends to have a lower bias with increasing estimation window as seen in Figure 4 (a), (c),
(d). This impact is especially significant when targeting an $\alpha$ below 5\%, but the impact is relatively negligible for any $N$ greater than 50.

\subsection{Effect of Number of Variables}

\begin{figure}[tb]
    \begin{tabular}{cc}
\includegraphics[width=0.45\textwidth]{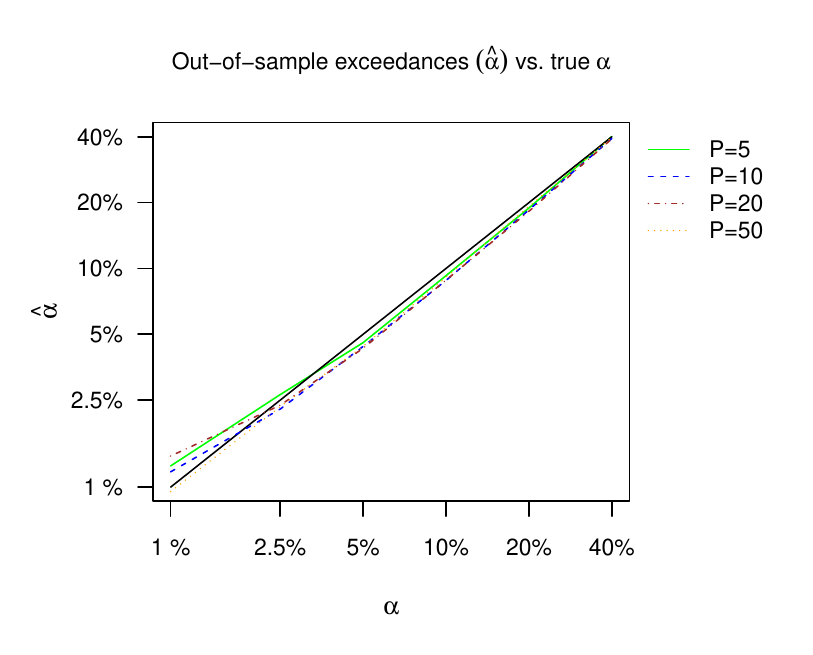}   
&
\includegraphics[width=0.45\textwidth]{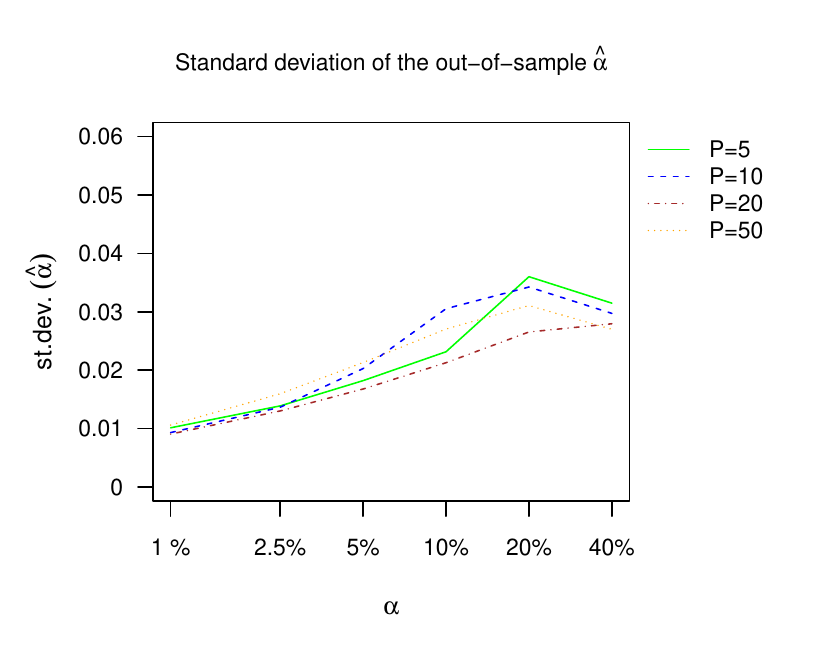} 
\\
(a) & (b)
\end{tabular}
\caption{Causal NECO VaR performance for different network sizes. (a) Log-scale comparison between $\hat{\alpha}$ achieved and $\alpha$ target for different numbers of instruments (p). (b) Standard deviation of the achieved $\hat{\alpha}$ at different values of the target $\alpha$ VaR level for different number of instruments (p). 
	}
\label{fig:sim-instruments}
\end{figure}

We then investigate the effect of the size of the contagion network, that is, the number of financial instruments and variables that we wish to analyse to estimate the causal structure. We simulate log-return financial networks with $p = 5, 10, 20, 50$ instruments. In the training simulation, N is set to 250, and the rest of the simulation is carried out as in the other simulations. From Figure \ref{fig:sim-instruments} we see that the impact of the size of the network is negligent, especially for the more common levels of $\alpha$ below 10\%. This finding is significant, suggesting that our causal NECO model maintains its efficacy in various network sizes, a crucial aspect for its application in various financial settings.

\subsection{Effect of Market Contagion}
\begin{figure}[]
	\begin{tabular}{cc}
\includegraphics[width=0.5\textwidth]{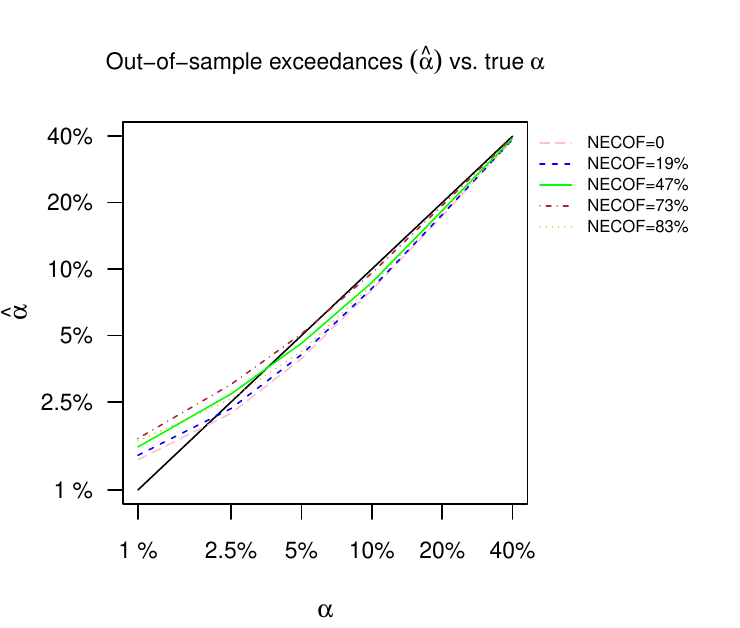} 
&
\includegraphics[width=0.5\textwidth]{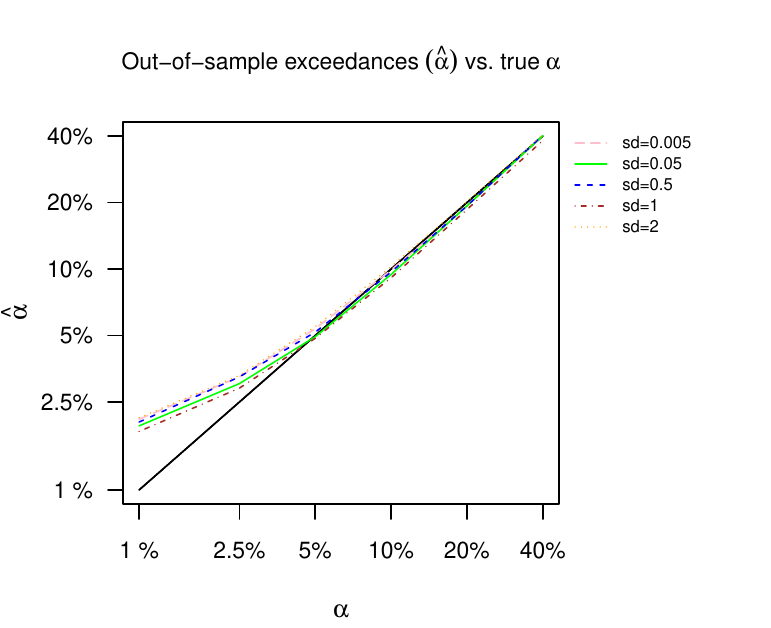} \\
(a) & (b)\end{tabular}
\caption{Causal NECO VaR perfomance comparison at a log-scale between target $\alpha$ and achieved $\hat{\alpha}$ VaR levels for (a) different levels of market contagion (NECOF) and (b) different levels of volatility ($\sigma$).}
\label{fig:sim-contagion-alphas}
\end{figure}

We compare the impact of market contagion, expressed in market NECOF, on the achieved $\hat{\alpha}$ VaR level. The study is carried out on a network of $p=10$ instruments and an estimation window of $N=250$. Market contagion is a function of a number of causal links in the financial network and the size of the causal contagion effects. We simulate different network structures with varying effect sizes and express the market contagion with the NECOF. Figure~\ref{fig:sim-contagion-alphas} (a) and Table~\ref{table:sim-contagion} show that there is no discernible trend in the efficiency of VaR estimation as the levels of contagion in the market increase. For small $\alpha$ below $5\%$, the method performs slightly better for lower contagion levels, suggesting that the choice of the length $N$ of the training window is especially crucial when targeting lower $\alpha$ levels in the case of extreme contagion levels. However, extreme levels of contagion would suggest a dense financial contagion network, which is not a typical situation for the financial market. From the left side of Table~\ref{table:sim-contagion} we see lower contagion levels. 

\begin{table}[tb]
\centering
\scalebox{0.8}{
\begin{tabular}{rrrrrr|rrrrr}
  \hline
  & \multicolumn{5}{c}{Contagion (NECOF)} & \multicolumn{5}{c}{Volatility (sd)} \\
 & 0\% & 19\% & 47\% & 73\% & 83\% 
 & 0.005 & 0.05 & 0.5 & 1 & 2\\ 
  \hline
 mean($\hat{\alpha}$) & 0.039 & 0.041 & 0.046 & 0.051 & 0.042 
   & 0.054 & 0.049 & 0.052 & 0.049 & 0.050\\ 
  sd($\hat{\alpha}$) & 0.015 & 0.017 & 0.022 & 0.021 & 0.020 
& 0.027 & 0.025 & 0.026 & 0.026 & 0.032\\ 
  LRuc.accept & 0.960 & 0.940 & 0.920 & 0.940 & 0.900 
  & 0.910 & 0.880 & 0.910 & 0.860 & 0.860\\ 
  LRcc.accept & 0.990 & 0.980 & 0.960 & 0.950 & 0.960 
  & 0.870 & 0.910 & 0.950 & 0.890 & 0.900\\ 
  DQ.accept & 0.880 & 0.860 & 0.820 & 0.810 & 0.880 
  & 0.850 & 0.890 & 0.830 & 0.860 & 0.840\\ 
  AE.mean & 0.786 & 0.819 & 0.922 & 1.017 & 0.849 
  & 1.074 & 0.988 & 1.036 & 0.972 & 1.098\\ 
  AE.sd & 0.306 & 0.329 & 0.436 & 0.427 & 0.399 
  & 0.538 & 0.495 & 0.531 & 0.525 & 0.643\\ 
  AD.mean & 0.024 & 0.026 & 0.029 & 0.035 & 0.035 
  & 0.000 & 0.000 & 0.000 & 0.000 & 0.000\\ 
  AD.max & 0.150 & 0.120 & 0.130 & 0.160 & 0.240 
  & 0.010 & 0.000 & 0.010 & 0.020 & 0.000\\ 
  CompareQL & 0.963 & 0.972 & 1.000 & 0.959 & 1.032
   & 0.983 & 1.034 & 1.000 & 1.028 & 0.986\\ 
   \hline
   ~~\\
\end{tabular}
}
\caption{Backtesting results for target $\alpha=5\%$ at different levels of market contagion (left) and different levels of volatility (right).}
\label{table:sim-contagion}
\end{table}

\subsection{Effect of Volatility}
Volatility, captured by the term $\Sigma$ in Equation~\ref{eqn:distribution}, is the stochasticity in the system.
As contagion also affects overall volatility, we keep the contagion levels fixed at 47\%. We simulate a system with $p=5$ instruments and keep the training window at $N=250$. We keep the shock to the system every 100 days but adapt the size to be proportional to the standard deviation.
The results for five different volatility levels $\sigma$ are shown in Figure~\ref{fig:sim-contagion-alphas} (b) and Table~\ref{tab:backtestsim}. The volatility change appears to have a negligible effect on the performance of causal NECO VaR. Regardless of volatility, the NECO VaR tends to be somewhat liberal at low VaR target values $\alpha$. This is mainly due to the external shocks included in the simulation. 

\subsection{Computational Time}
\begin{figure}[]
	\centering
\includegraphics[width=0.5\textwidth]{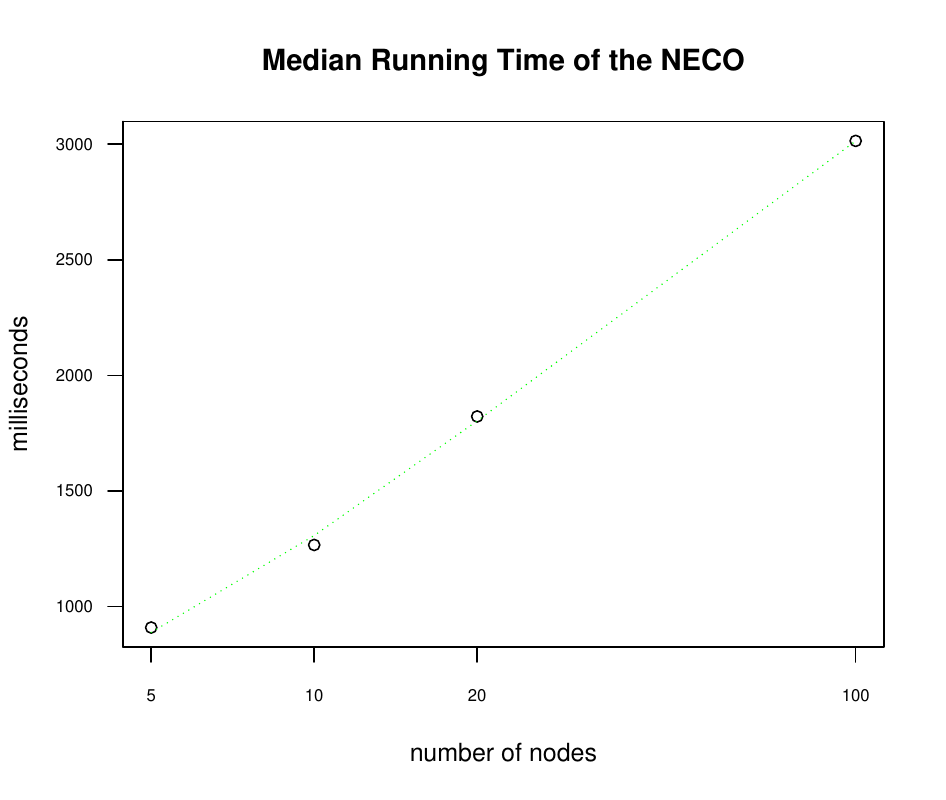}
\caption{Average computation time in milliseconds across 100 simulations of the causal NECO network identification and estimation as a function of the number of financial instruments (nodes).}
\label{fig:sim-time}
\end{figure}

The causal inference approach is well adapted to sparse networks that have relatively few causal links \citep{le2016fast}. This seems to be the case for financial networks, as there are usually clear pathways through which contagion flows \citep{bardoscia2021physics}. Figure~\ref{fig:sim-time} shows that for such sparse networks, the computational time is low, even for a high number of financial instruments. It is pertinent to note that in scenarios involving extremely dense networks or an exceptionally high number of nodes, the PC-stable algorithm, which underpins our methodology, may encounter convergence challenges \citep{kalisch2007estimating}. 

\section{Measuring Risk on Forex}\label{forex}
Forex is a very liquid and important financial market, trading \$ 6.6 trillion per day \citep{bank2019triennial}. 90\% of these transactions involve the US dollar (USD). In this study, we analyse the Forex market of 20 exchange rates over the USD, selected to provide a representative overview of the most commonly traded currencies, as well as some less common ones to showcase different challenges in risk estimation. Given that we consider liquid assets, we will consider a VaR with a 1-day holding period.

Table \ref{table:SummaryForex2} shows the summary statistics of the daily log-retuns on 20 considered exchange rates over the USD for the years 2000-2021, as published by the Federal Reserve of New York.  Most of the currencies have considerable fat tails in their log-returns. Most currencies show significant deviations from normality, as evidenced by the Jarque-Bera test statistics, which are notably high across the board. Most of the currencies have considerable fat tails and asymmetry in their log-returns. NECOF values, which indicate the degree of the contagion effect that impacts the specific currency, vary considerably, suggesting various levels of interconnection and risk exposure.


\begin{table}[tb]
\centering
\scalebox{0.7}{
\begin{tabular}{rrrrrrrrrr}
  \hline
 &$\NECOFest$ & Minimum & Median & Mean & Maximum & StDev & Skewness & Kurtosis & Jarque Bera \\ 
  \hline
AUD &0\% & -0.0771 & -0.0003 & -0.0001 & 0.0822 & 0.0080 & 0.6197 & 12.0299 & 31085\\ 
  EUR &0\% & -0.0463 & 0 & -0.0001 & 0.0300 & 0.0059 & -0.0557 & 2.5411 & 1375\\ 
  NZD &63.9\%& -0.0593 & -0.0003 & -0.0001 & 0.0618 & 0.0082 & 0.3770 & 4.8517 & 5124\\ 
  GBP &20.2\%& -0.0443 & -0.0001 & 0 & 0.0817 & 0.0060 & 0.7047 & 10.8604 & 25491\\ 
  BRL &5.8\%& -0.0967 & 0 & 0.0002 & 0.0867 & 0.0105 & -0.0040 & 8.1012 & 13949\\ 
  CAD &37.6\%& -0.0507 & -0.0001 & -0 & 0.0381 & 0.0057 & -0.0669 & 5.4645 & 6350\\ 
  DKK &98.5\%& -0.0580 & -0.0001 & -0.0001 & 0.0494 & 0.0060 & -0.1284 & 4.7956 & 4902\\ 
  HKD &0.3\%& -0.0045 & 0 & -0 & 0.0033 & 0.0003 & -1.2257 & 25.3347 & 137697\\ 
  INR &1.3\%& -0.0376 & 0 & 0.0001 & 0.0394 & 0.0045 & 0.1980 & 9.9731 & 21174\\ 
  JPY &6.2\%& -0.0522 & 0.0001 & -0 & 0.0334 & 0.0062 & -0.3144 & 4.4245 & 4245\\ 
  KRW &13.3\%& -0.1322 & -0.0001 & -0 & 0.1014 & 0.0068 & -0.5511 & 49.7260 & 525804\\ 
  MXN &37.8\%& -0.0596 & -0.0001 & 0.0001 & 0.0811 & 0.0072 & 0.7485 & 11.0121 & 26250\\ 
  NOK &40.6\%& -0.0644 & -0.0002 & -0 & 0.0612 & 0.0078 & 0.2380 & 4.8194 & 4985\\ 
  SEK &63.3\%& -0.0530 & -0 & -0 & 0.0547 & 0.0074 & -0.0482 & 3.9502 & 3319\\ 
  ZAR &42.4\%& -0.0916 & -0.0002 & 0.0001 & 0.0843 & 0.0109 & 0.2626 & 4.2637 & 3922\\ 
  SGD &59.5\%& -0.0238 & -0.0001 & -0.0001 & 0.0269 & 0.0033 & 0.0313 & 4.9083 & 5121\\ 
  LKR &0.4\%& -0.0339 & 0 & 0.0002 & 0.0641 & 0.0029 & 2.5329 & 76.6587 & 1254466\\ 
  CHF &65.6\%& -0.1302 & 0 & -0.0001 & 0.0889 & 0.0067 & -1.1545 & 36.8482 & 289720\\ 
  TWD &39.5\%& -0.0342 & 0 & -0 & 0.0248 & 0.0030 & -0.3858 & 9.8420 & 20714\\ 
  THB &37.9\%& -0.0353 & 0 & -0.0001 & 0.0447 & 0.0037 & 0.1609 & 12.2860 & 32104\\ 
   \hline
\end{tabular}
}
\caption{Overview of the summary statistics for the dataset of log-returns on individual 20 exchange rates over the USD, for the period January 2000 to April 2021. The higher the Jarque Bera test statistic the less likely the data are normally distributed --- all of the statistics have a p-value of 0. Each sample is 5101 observations long with no values missing.}
\label{table:SummaryForex2}
\end{table}

\subsection{Fitting Causal Network Contagion on Forex}
\begin{figure}[]
	\begin{center}
		\includegraphics[width=0.5\textwidth]{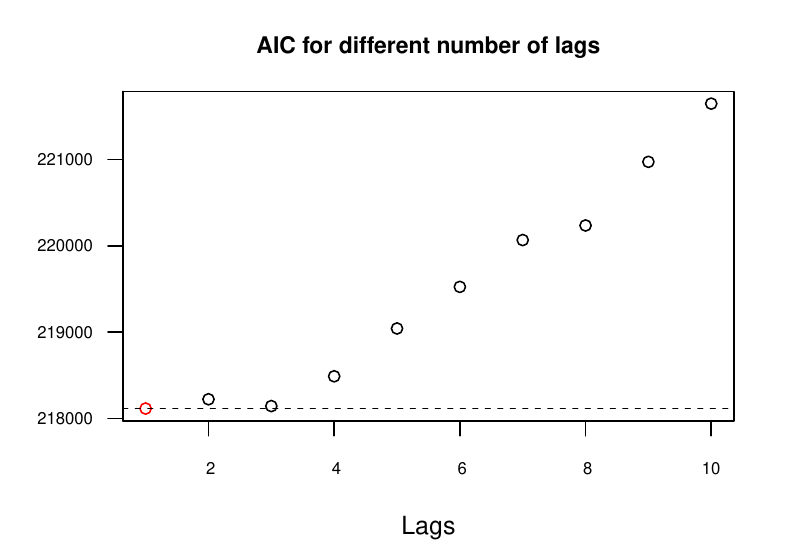}
		\caption{Overall AIC for each number of lags on the whole of the Forex dataset considered.} 
		\label{fig:AIC}
	\end{center}
\end{figure}

We consider causal networks with different values of the lag $L$ in Equation~\ref{eqn:commented}. In order to choose the best number of lags, we make use of the Akaike information criterion, given as $\mbox{AIC}(\ell) = 2k(\ell) - 2l_\ell(D_X)$, where $l$ is the log-likelihood, and $k(\ell)$ is the number of all non-zero autoregressive parameters $A$ and contagion coefficients $B$. Figure~\ref{fig:AIC} shows the value of the AIC for different lags. The minimum is obtained at lag $\ell=1$, which is used in the estimated NECO model. 

Given the presence of fat tails in the log-return, we use a copula transformation, as described in section~\ref{sec:necovar}. We fit a marginal distribution $F_i$ for the log-returns for each of the 20 currencies. We fit the causal network to the transformed log-returns as explained in Section \ref{estimation}. Finally, we obtain the 1-day ahead causal NECO VaR values via Equation~\ref{eqn:neco-var-general}.



\subsection{Results and Backtesting}

We will consider the Value at Risk for each individual currency rate at the $\alpha$ level of 5\%. We consider a training window of $N=250$ trading days. We will compare the performance of the Causal NECO approach to the more established methods HIST, VarCovar, GARCH, and FHS-GARCH with the backtesting measures from Section~\ref{compare-def} applied to 100 out-of-sample VaR predictions, for each of the 20 currencies during 20 non-overlapping periods between 2000 and 2021.

Table~\ref{table:BacktestForex5} shows that the causal NECO VaR beats all other methods in most categories with a much higher acceptance rate for all tests. We also see a much lower standard deviation of the estimated $\hat{\alpha}$, in line with a robust invariant prediction. Given that, at least during calm periods, the overall volatility of the Forex is quite low, the losses are not extreme for any of the methods. FHS seems to be the best competition for causal NECO VaR at this level.

\begin{table}[tb]
\centering
\small 
\begin{tabular}{rrrrrr}
  \hline
 & Causal-NECO & VarCovar & HIST & GARCH & FHS \\ 
  \hline
 mean$(\hat{\alpha})$  & \textbf{0.0511 }& 0.0597 & 0.0594 & 0.0549 & 0.0574 \\ 
   st.dev$(\hat{\alpha})$  & \textbf{0.0202 }& 0.0603 & 0.0392 & 0.0338 & 0.0318 \\ 
  LRuc.accept & \textbf{0.9500} & 0.8100 & 0.6700 & 0.7600 & 0.7800 \\ 
  LRcc.accept & \textbf{0.9600} & 0.9000 & 0.8000 & 0.8700 & 0.8800 \\ 
  DQ.accept & \textbf{0.8500} & 0.8000 & 0.7800 & 0.8200 &\textbf{ 0.8500} \\ 
  AE.mean & 1.0220 & 1.1060 & 1.0995 &\textbf{ 1.0085} & 1.0595 \\ 
  AE.sd & \textbf{0.4046} & 1.2171 & 0.7928 & 0.6818 & 0.6416 \\ 
  AD.mean & 0.0032 & 0.0031 & 0.0032 &\textbf{ 0.0030 }& \textbf{0.0030} \\ 
  AD.max & 0.1200 & 0.1200 & 0.1200 & 0.1200 & 0.1200 \\
  CompareQL & 1.0000 & 0.9804 & 1.0247 & 0.9934 & \textbf{0.9801} \\ 
   \hline
\end{tabular}
\normalsize
\caption{Backtesting of the VaR out-of-sample predictions in the Forex market (2000-2021) for the different methods at target $\alpha=5\%$. The best results are presented in bold.}
\label{table:BacktestForex5}
\end{table}

\begin{figure}[]
    \tabcolsep = 1.0pt
    \begin{tabular}{cc}
\includegraphics[width=0.4\textwidth]{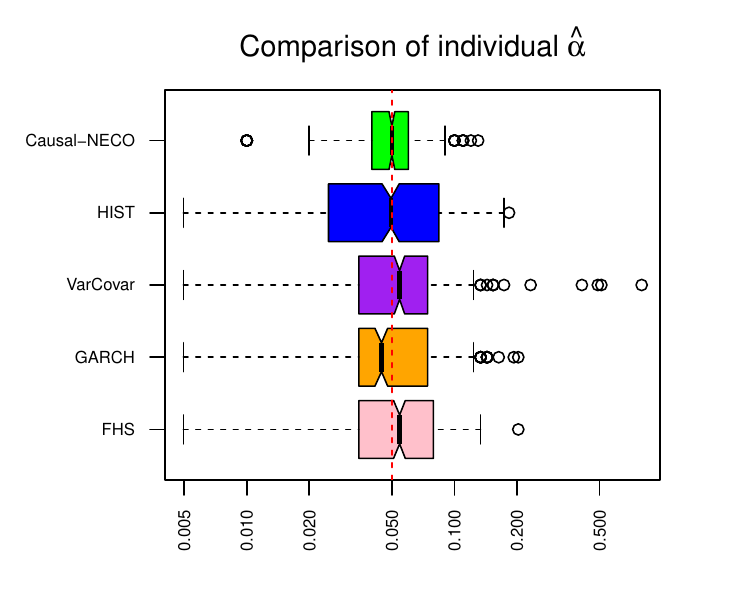}   
&
\includegraphics[width=0.5\textwidth]{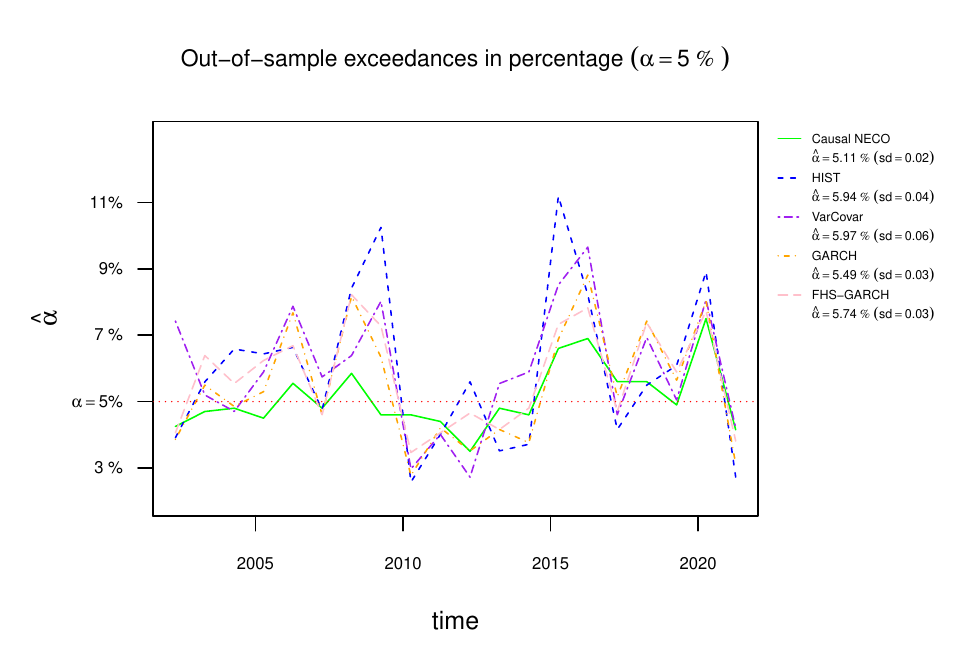} 
\\
(a) & (b)
\end{tabular}
\caption{(a) Overall fraction of exceedances $\hat{\alpha}$ within the out-of-sample window of 100 days for each Value-at-Risk model across the Forex market (2000-2021). (b) Fraction of the exceedances over time.}
\label{fig:forex-compare-5}
\end{figure}

Figure \ref{fig:forex-compare-5} (a) shows the distribution of actual exceedances for all currencies and periods. Causal-NECO is the most centered around the target value of 5\%. 
Figure \ref{fig:forex-compare-5} (b) shows how these exceedances change over time.  The Causal-NECO method seems to be able to adapt properly to the changing conditions of the underlying network. Unlike the other methods, it is barely affected by the various financial crises in this period and achieves the nominal 5\% level throughout the evaluation period. Even as other methods seem to underestimate the risk during peaks such as the 2008 global financial crisis, and move all together, Causal-NECO keeps tight on target.

\begin{figure}[]
\includegraphics[width=\textwidth]{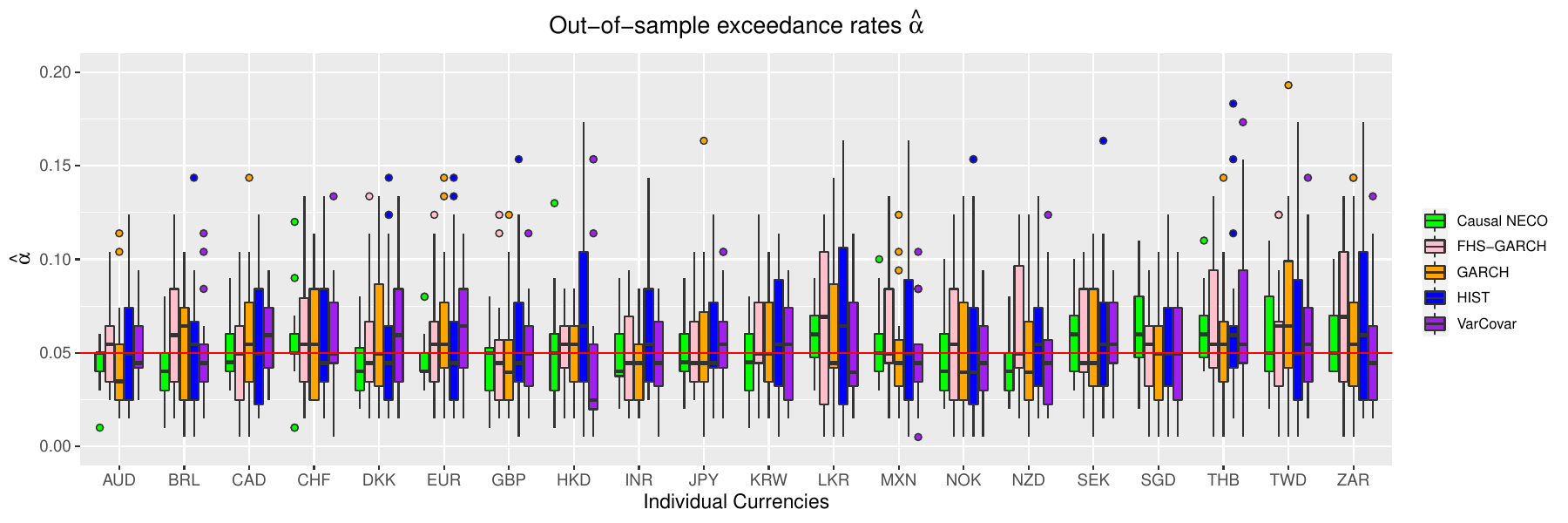}
\caption{Number of actual exceedances within the out-of-sample window of 100 days for each Value-at-Risk model, expressed in terms of the $\hat{\alpha}$ computed for each period and shown for each currency separately. The red line shows the target of $\alpha=5\%$.}
\label{fig:Boxplot_Forex-currencies}
\end{figure}

In addition, we consider each of the 20 currencies individually. Figure~\ref{fig:Boxplot_Forex-currencies} shows the individual boxplots of the exceedance fractions for each of the given methods. The causal NECO performance is consistently less variable and more centred around the 5\% level. 

\section{Conclusion}\label{conclusion2}

This paper introduces an innovative way to eliminate spurious correlations in risk management decisions. It lays the groundwork for using causal inference in finance showing how practical advantages can be obtained by inferring the underlying causal system from available financial data. The causal NECO VaR is easy to compute and offers a robust and competitive addition to standard approaches. Volatility is not modelled directly but is addressed through the contagion effect and, in part, through the copula transformation, offering an alternative view on modelling volatility.
As a way to deal with the non-normality of financial data, we use a Gaussian copula transformation improving upon its historical applications and addressing some of the limitations highlighted in past financial crises. In our case, it is also important to note that we do not treat the Gaussian copula as a way to model risk but rather to improve the estimation of our causal contagion network. 

The Causal NECO VaR model presents numerous opportunities for expansion. Future research could attempt to account for a wider class of distributions using a Student-t copula instead of a Gaussian copula and time-varying errors with an added GARCH component \citep{jondeau2006copula}. However, both extensions are beyond the scope of this paper and could theoretically be incorporated without affecting the underlying causal structure of our model. 
%
Future research could explore the effectiveness of the causal approach across various financial instruments. Our model allows for the inclusion of any measurable confounder, which could be used to assess the impact of factors such as interest rates, inflation, or stock market returns on Forex or other contagion networks. In cases involving unmeasured confounders and latent variables, the Fast Causal Inference Algorithm (FCI) \citep{spirtes1995causal,spirtes2000causation} presents a suitable alternative. In addition, incorporating nonparametric methods into our model, as discussed in the recent literature \citep{li2020nonparametric}, would further enhance its adaptability.

In conclusion, the Causal NECO VaR model marks a significant advancement in financial risk analysis. Its ability to incorporate causal relationships into risk assessment not only broadens our understanding of market dynamics but also paves the way for more resilient financial management strategies in an ever-evolving economic landscape.



\myacknowledgments
\bibliographystyle{chicago}
\bibliography{biblio}

\end{document}